\documentclass[twocolumn,showpacs,aps,prl,superscriptaddress]{revtex4}

\usepackage{graphicx}
\usepackage{amsmath}

\begin{document}

\title{Valley splitting of Si/Si$_{1-x}$Ge$_x$ heterostructures in tilted magnetic fields}

\author{K. Lai}
\affiliation{Department of Electrical Engineering, Princeton
University, Princeton, New Jersey 08544}
\author{T.M. Lu}
\affiliation{Department of Electrical Engineering, Princeton
University, Princeton, New Jersey 08544}
\author{W. Pan}
\affiliation{Sandia National Laboratories, Albuquerque, NM
87185}
\author{D.C. Tsui} \affiliation{Department of Electrical
Engineering, Princeton University, Princeton, New Jersey 08544}
\author{S. Lyon}
\affiliation{Department of Electrical Engineering, Princeton
University, Princeton, New Jersey 08544}
\author{J. Liu}
\affiliation{Department of Material Science and Engineering,
UCLA, Los Angeles, CA 90095}
\author{Y.H. Xie}
\affiliation{Department of Material Science and Engineering,
UCLA, Los Angeles, CA 90095}
\author{M. M\"uhlberger}
\affiliation{Institut f\"ur Halbleiterphysik, Universit\"at Linz, A-4040 Linz, Austria}
\author{F. Sch\"affler}
\affiliation{Institut f\"ur Halbleiterphysik, Universit\"at Linz, A-4040 Linz, Austria}

\date{\today}

\begin{abstract}

We have investigated the valley splitting of two-dimensional
electrons in high quality Si/Si$_{1-x}$Ge$_x$ heterostructures under
tilted magnetic fields. For all the samples in our study, the valley
splitting at filling factor $\nu=3$ ($\Delta_3$) is significantly
different before and after the coincidence angle, at which energy
levels cross at the Fermi level. On both sides of the coincidence, a
linear density dependence of $\Delta_3$ on the electron density was
observed, while the slope of these two configurations differs by
more than a factor of two. We argue that screening of the Coulomb
interaction from the low-lying filled levels, which also explains
the observed spin-dependent resistivity, is responsible for the
large difference of $\Delta_3$ before and after the coincidence.

\end{abstract}
\pacs{73.43.Fg,73.21.-b}
\maketitle

The study on the valley splitting of the two-dimensional electron
gas (2DEG) confined in (001) Si surface has been highlighted by
recent research effort on Si-based quantum computation\cite {xiao}.
For a Si 2DEG, only the two out-of-plane valleys are relevant since
the other four in-plane valleys are lifted from the conduction band
edge. To realize a functional Si quantum computer using spins as
quantum bits, a large valley splitting that lifts the remaining
two-fold degeneracy is desirable since the existence of two
degenerate states associated with the $\pm$k$\rm_z$ valleys is
believed to be a potential source of spin decoherence \cite {xiao}.
In the single-particle picture, theories \cite {ohkawa, sham, ando}
in the early period of the 2D physics proposed that the surface
electric field in the presence of 2D interface breaks the symmetry
of these two valleys, resulting in an energy splitting proportional
to the carrier density. The understanding of the valley splitting in
real Si systems, however, is not a trivial task and requires much
beyond such non-interacting band picture. In fact, the many-body
effect \cite {ohkawa, ando} was speculated to account for the
enhancement over the bare valley splitting under strong magnetic (B)
fields, while a detailed calculation is not yet available.

Experimental research on the valley splitting, on the other hand,
was conducted mainly on the Si metal-oxide-semiconductor
field-effect transistors (MOSFETs), in which the disorder effect is
strong and direct measurement of the valley splitting proves to be
difficult \cite{nicholas, pudalov}. More than a decade ago, the
introduction of the graded buffer scheme significantly improved the
sample quality of the Si/SiGe heterostructures \cite{schaffler}. To
date, the valley splitting has been studied by various experimental
techniques, including thermal activation \cite{weitz}, tilted field
magnetotransport \cite{koester, schumacher}, magnetocapacitance
\cite{khrapai}, microwave photoconductivity \cite{goswami} and
magnetization \cite{wilde}. However, as pointed out by Wilde $et$
$al$. in Ref. \cite{wilde}, results reported by different groups are
ambiguous and inconsistent with previous band calculations. The
nature of this valley splitting, especially its behavior under
strong B-fields, stays as an unsettled problem.

Of the various methods used to study the valley splitting, tilted
field magnetotransport, also known as the coincidence method
\cite{fang}, is frequently utilized. In a B-field tilted by an angle
$\theta$ with respect to the 2D plane, the ratio of the cyclotron
energy E$\rm_C$ = $\hbar$$\omega\rm_C$ =
$\hbar$eB$_{\perp}$/m$^\ast$, where B$_{\perp}$ is the perpendicular
field and m$^\ast$ the effective mass, to the Zeeman energy E$_Z$ =
g$^\ast$$\mu\rm_B$B$\rm_{tot}$, where g$^\ast$ is the effective g-factor,
$\mu\rm_B$ the Bohr magneton and B$\rm_{tot}$ the total field, can 
be continuously tuned by adjusting $\theta$ = cos$^{-1}$(B$_\perp$/B$\rm_{tot}$). 
In particular, the so-called coincidence happens when the energy levels
from different Landau levels (LLs) are aligned at the Fermi level.
In a recent experiment \cite{lai}, the inter-valley energy gaps at
the odd-integer quantum Hall (QH) states were studied and found to
rise rapidly towards the coincidence. In this work, we show that the
anomalous rise was not observed in the even-integer QH states, whose
energy gaps close as $\theta$ approaches the single-particle
degenerate points. For all the samples in our study, the $\nu=3$
valley splitting before the coincidence follows a linear density
dependence that extrapolates to about -0.4K at zero density, which
is probably due to level broadening. The $\nu=3$ gap after the
coincidence also depends linearly on density, while the slope
increases by more than a factor of two. We argue that screening of
the Coulomb interaction from the low-lying filled levels, which also
explains the observed spin-dependent resistivity, is responsible for
the change of the observed $\nu=3$ gaps on different sides of the
coincidence.

The specimens in our study are modulation-doped n-type Si/SiGe
heterostructures grown by molecular-beam epitaxy. Important sample
parameters, such as the electron density ($n$), mobility ($\mu$) and
width of the quantum well ($W$), are listed in Table. 1. For the
samples labeled as LJxxx, relaxed Si$_{0.8}$Ge$_{0.2}$ buffers
provided by Advanced Micro Devices (AMD) were used as substrates,
followed by a 1 $\mu$m Si$_{0.8}$Ge$_{0.2}$ buffer layer prior to
the growth of the strained Si channel. On top of the Si quantum
well, a 20nm Si$_{0.8}$Ge$_{0.2}$ spacer, a delta-doped Sb layer, a
25nm Si$_{0.8}$Ge$_{0.2}$ cap, and a 4nm Si cap layer are
subsequently grown. The carrier density is controlled by the amount
of Sb dopants. The high mobility sample labeled as 1317 is the same
specimen as that used in Ref. \cite{lai} and its density and
mobility can be tuned by controlling the dose of low temperature
illumination by a light-emitting diode (LED).
\\
\\
\\
\begin{tabular}{|c|c|c|c|c|}\hline
Sample & $n$(10$^{11}$cm$^{-2}$) & $\mu$(m$^2$/Vs) & $W$(nm) & Illumination \\
\hline LJ122 & 3.1 & 6.3 & 10 & No\\
\hline LJ126 & 2.3 & 9.8 & 10 & Saturated\\
\hline LJ127 & 2.1 & 8.7 & 10 & Saturated\\
\hline LJ139 & 1.7 & 12 & 20 & Saturated\\
\hline 1317-I & 1.4 & 19 & 15 & No\\
\hline 1317-II & 1.8 & 22 & 15 & Unsaturated\\
\hline 1317-III & 2.4 & 25 & 15 & Saturated\\
\hline
\end{tabular}
\\
\\
{\small Table 1. List of sample parameters. The density, mobility
and width of the quantum well are shown, together with the dose of
illumination.}
\\

Magnetotransport measurements were performed in the 18/20T
superconducting magnet in the National High Magnetic Field
Laboratory (NHMFL) in Tallahassee, FL. Samples were sitting in a
rotating stage at the dilution refrigerator with a base temperature
T$\rm_{base}$ = 20mK. Standard low frequency (5$\sim$13Hz) lock-in
techniques were used to measure the diagonal resistivity
$\rho\rm_{xx}$ and the Hall resistivity $\rho\rm_{xy}$.

\begin{figure}[!t]
\begin{center}
\includegraphics[width=3.2in,trim=0.3in 0.5in 0.2in 0.2in]{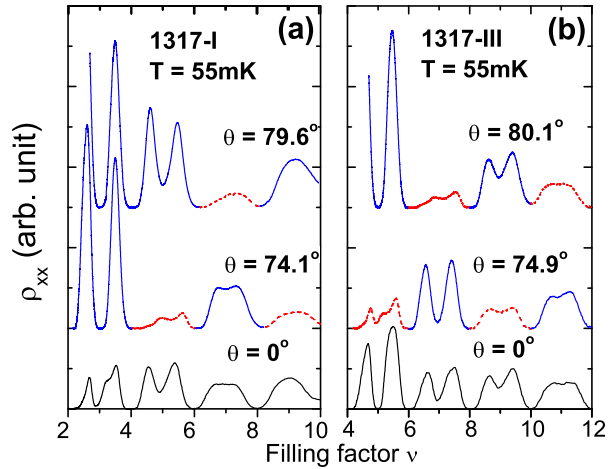}
\end{center}
\caption{\label{1} Magnetoresistivity $\rho\rm_{xx}$ as a function
of the filling factor for samples (a) 1317-I and (b) 1317-III at
selected tilt angles at T = 55mK. From bottom to top, the system is
before the 1st coincidence ($\theta$ = 0$^o$), between the 1st and
2nd coincidences ($\theta$ $\sim$ 74$^o$), and after the 2nd
coincidence ($\theta$ $\sim$ 80$^o$). After the 1st coincidence, the
overall amplitude of the $\rho\rm_{xx}$ is generally higher when
electrons in the Fermi level have up-spins (solid blue curves) and
lower for down-spins (dashed red curves).}
\end{figure}

In Fig. 1, we show the $\rho\rm_{xx}$ traces as a function of the
filling factor ($\nu$) at several tilt angles for samples (a) 1317-I
and (b) 1317-III. The odd-integer QH states $\nu$ = 3, 5... are
associated with energy gaps opened by the valley splitting. The
three tilt angles were chosen so that from the bottom to the top
traces, 1/cos$\theta$ = 1 (before the 1st coincidence), $\sim$ 3.7
(between the 1st and 2nd coincidences) and $\sim$ 5.6 (after the 2nd
coincidence), respectively. We will return to the tilt-field data
later in the discussion.

Fig. 2a shows a schematic of the tilted-field energy diagram of a Si
2DEG. The LL (N), spin ($\uparrow$ or $\downarrow$) and valley (+ or
--) indices are indicated in the plot. Since $\Delta\rm_v$ is shown
to be independent of the parallel field \cite{weitz, lai}, the two
valley states originated from each spin level are parallel to each
other in the diagram. In this independent-electron picture, the
levels are not affected as they cross each other, and the energy gap
of individual QH states closes at certain tilt angles, or
coincidence angles. Since in a Si 2DEG, $\Delta\rm_v$ is usually
much smaller than E$\rm_Z$ and E$\rm_C$, we adopt the conventional
notation that the $j$th order coincidence occurs when E$\rm_Z$ /
E$\rm_C$ roughly equals an integer number $j$. In Fig. 2b and 2c,
the energy gaps, obtained by fitting $\rho_{xx}$$\propto$
exp(-$\Delta_3$/2k$_B$T) in the thermal activation regime, at $\nu=
4$ and 6 in sample 1317-I are shown as a function of 1/cos$\theta$
or B$\rm_{tot}$/B$_{\perp}$. When $\theta$ is away from the
coincidences, the gaps at $\nu=4$ and 6 vary linearly with respect
to 1/cos$\theta$ with a slope corresponding to g$^\ast$ = 2,
consistent with the independent-electron model. On the other hand,
the even-integer energy gaps drop suddenly towards the coincidence
angles at which the single-particle gap closes, e.g., 1/cos$\theta$
$\sim$ 2.5 (1st coincidence) for $\nu$ = 4 and 1/cos$\theta$ $\sim$
4.5 (2nd coincidence) for $\nu$ = 6, as can be seen in Fig. 2b and
2c. This sudden drop of activation gap towards the degenerate points
was observed in a wide GaAs/AlGaAs quantum well and explained within
the framework of quantum Hall ferromagnetism \cite{muraki}.

\begin{figure}[!t]
\begin{center}
\includegraphics[width=3.2in,trim=0.3in 0.5in 0.2in 0.2in]{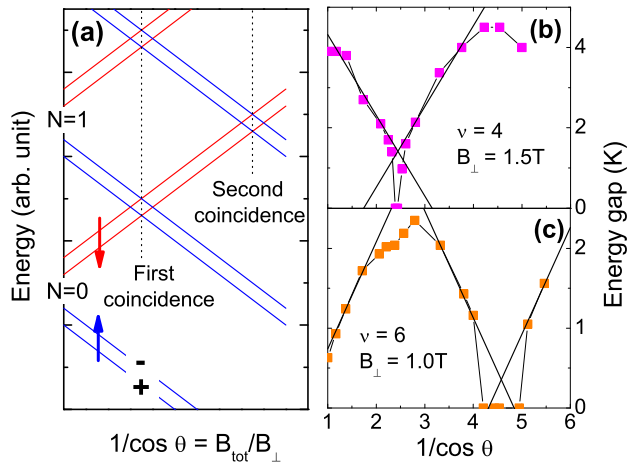}
\end{center}
\caption{\label{2} (a) Schematic of the LL fan diagram in tilted
B-fields. The LL (N), spin ($\uparrow$ or$\downarrow$) and valley (+
or --) indices are indicated for each level. The positions of the
1st and 2nd coincidences are indicated. (b) Measured energy gaps at
$\nu$ = 4 (B$_{\perp}$ = 1.5T) and (c) $\nu$ = 6 (B$_{\perp}$ =
1.0T) of sample 1317-I as a function of 1/cos$\theta$ or
B$\rm_{tot}$/B$_{\perp}$. The solid lines correspond to g$^{\ast}$ =
2.}
\end{figure}

In contrast to the well-behaved even-integer QH states, the energy
gap of the $\nu$ = 3 state ($\Delta_3$) exhibits an anomalous rise
towards the coincidence, as shown in the inset of Fig. 3, a
phenomenon previously reported in Ref. \cite{lai}. We emphasize here
that such an anomaly was observed in all the samples investigated in
this study, in spite of the considerable difference in the sample
structure and mobility. Out of the coincidence region, the
activation energy is indeed independent of the parallel field
component, while it differs by about a factor of 3 (0.8K vs. 2.1K)
on different sides of the coincidence. Referring to the level
diagram in Fig. 2a, we label the valley splitting as
$\Delta_3$(N=0,$\downarrow$) and $\Delta_3$(N=1,$\uparrow$) before
and after the coincidence, respectively.

In Fig. 3, we plot the measured $\Delta_3$(N=0,$\downarrow$) and
$\Delta_3$(N=1,$\uparrow$) gaps for all 7 samples as a function of
the carrier density. The band calculation of valley splitting in a
Si 2DEG \cite{ohkawa, sham, ando} based on the effective-mass
approximation, showing a linear dependence $\Delta\rm_v$ (K) $\sim$
0.17$n$ (10$^{11}$cm$^{-2}$) at B = 0, is also plotted (solid line)
for comparison. Despite some scattering in the data, the measured
$\Delta_3$(N=0,$\downarrow$) gaps essentially fall on a straight
line that extrapolates to -0.4$\pm$0.2K at zero density. We note
that this energy of -0.4K is within the order of the
sample-dependent disorder broadening ($\Gamma$ $\sim$ $\hbar$/$\tau$
= $\hbar$e/m$^\ast$$\mu$ , where $\tau$ is the transport scattering
time), which lies between 0.3K and 1.1K in our samples.
Interestingly, the detailed sample structure, e.g., the well width
$W$, seems less important here. The $\Delta_3$(N=1,$\uparrow$) gaps
of the same set of samples also fall onto a line extrapolating to
-0.7$\pm$0.3K at $n$ = 0, again within the order of level
broadening. On the other hand, the slope of the linear density
dependence differs by more than a factor of 2 (0.5K vs. 1.4K per
10$^{11}$cm$^{-2}$) before and after the coincidence. And both are
significantly higher than that of the band calculation at B=0.

\begin{figure}[!t]
\begin{center}
\includegraphics[width=3.2in,trim=0.3in 0.5in 0.2in 0.2in]{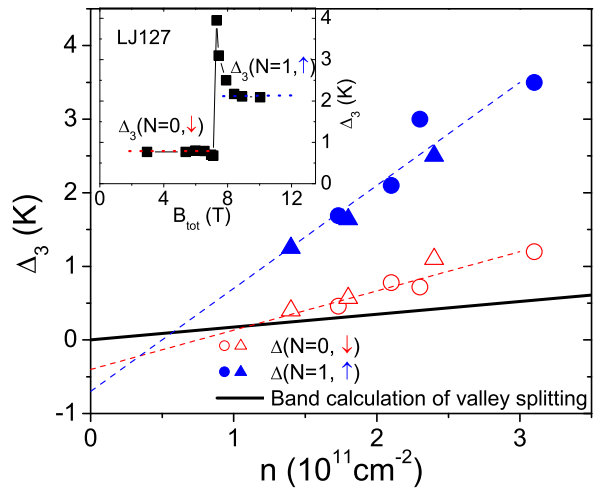}
\end{center}
\caption{\label{3} Density dependence of the valley splitting at
$\nu$ = 3. The empty symbols (triangles for samples 1317 and circles
for LJxxx) stand for $\Delta_3$(N=0,$\downarrow$) and the filled
symbols for $\Delta_3$(N=1,$\uparrow$). Dashed lines are linear fits
to the data and extrapolate to finite values at zero density. The
solid line shows the band calculation of valley splitting in Ref.
\cite{ohkawa}. The inset shows the $\Delta_3$ gap of sample LJ127 as
a function of B$\rm_{tot}$. The coincidence occurs around
B$\rm_{tot}$ = 7T.}
\end{figure}

The linear density dependence of the valley gaps and strong
enhancement over the bare valley splitting were recently reported in
a Si-MOSFET system using magnetocapacitance method \cite{khrapai}.
The authors pointed out that the electron-electron (e-e)
interaction, especially the exchange interaction, is likely to
account for the observed large valley gaps. In order to shed some
light to the apparent large difference between the
$\Delta_3$(N=0,$\downarrow$) and $\Delta_3$(N=1,$\uparrow$) gaps, we
scrutinize the many-body effect for the two configurations of $\nu$
= 3, shown in Fig. 4. For the relevant perpendicular B-fields in
this work, the e-e interaction energy E$\rm_{e-e}$ $\sim$
e$^2$/4$\pi\epsilon$$l\rm_B$ ($l\rm_B$ =
($\hbar$/eB$_{\perp}$)$^{-1/2}$ is the magnetic length) is larger
than the LL spacing so mixing between different LLs has to be taken
into account. Consequently, we explicitly include the lower two
filled levels (N=0, $\uparrow$, $\pm$), which are kept intact for
all tilt angles, into the analysis. Before the coincidence,
electrons in these two low-lying levels have the same LL but
opposite spin indices comparing to the ones near the Fermi level
(E$\rm_F$). Since the Pauli exclusion principle does not prevent the
opposite spins from approaching each other, these low-lying
electrons can come close to the electrons at E$\rm_F$ and strongly
screen the Coulomb interaction. The enhancement of the $\nu$ = 3 gap
due to the electron-electron interaction is thus much reduced and
the gap is close to the bare value at this LL. On the other side of
the coincidence, however, such screening is much less effective.
First, the electrons near E$\rm_F$ are from the N=1 LL and their
wave function is different from the N=0 levels. The off-diagonal
matrix element of this Coulomb energy between the two different LLs
should be considerably smaller than that from the same LL. Second,
even in the presence of LL mixing effect, the exclusion principle
limits the screening between the same up-spin levels. As a result,
the $\Delta_3$(N=1,$\uparrow$) gap is greatly enhanced over the bare
valley splitting. We nevertheless emphasize here that in the last
few LLs, the shape of the wave function is completely different from
the plane wave at B = 0. So even the bare valley splitting here
could be different from the results obtained by Ohkawa and Uemura
\cite{ohkawa}, who only consider high LLs by using simple average
over the in-plane k-vector.

\begin{figure}[!t]
\begin{center}
\includegraphics[width=2.8in,trim=0.5in 1.0in 0.5in 0.8in]{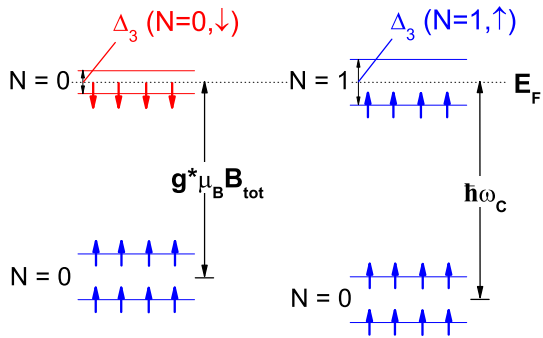}
\end{center}
\caption{\label{4} Level diagram at $\nu$ = 3 before (left) and
after (right) the coincidence. E$\rm_F$ resides in the gap between
the lowest empty levels and the top filled levels. The level
occupation, as well as the spin orientation, is indicated in the
plot. Before the coincidence, the low-lying (N=0,$\uparrow$,$\pm$)
electrons, separated by E$\rm_Z$ = g$^{\ast}$$\mu_B$B$\rm_{tot}$
from the Fermi level, strongly screen the Coulomb interaction for
electrons near E$\rm_F$, resulting in a less enhanced
$\Delta_3$(N=0,$\downarrow$) over the bare valley splitting. The
same screening, on the other hand, is less effective from the
like-spin charges in a different LL, giving a large
$\Delta_3$(N=1,$\uparrow$).}
\end{figure}

Finally, we note that the spin-dependent resistivity, first reported
by Vakili $et$ $al$. \cite{vakili} and successfully explained by
screening from the filled LLs, is also observed in our samples. In
Fig. 1, after the 1st coincidence, the overall $\rho\rm_{xx}$
amplitude is lower (dashed red curves) when the spins at the Fermi
level orient opposite to the majority up-spins in the system and
higher when the two are aligned (solid blue curves), which was
attributed to screening from the low-lying filled LLs. Due to the
exclusion principle, electrons with same spins cannot approach each
other to effectively screen the disorder potential, resulting in a
higher $\rho\rm_{xx}$ comparing to the opposite case. Interestingly,
the same alternating pattern is also observed in the strengths of
the odd-integer valley states.

In summary, we have carried out a titled field study of the Si/SiGe
heterostructures and measured the energy gaps of integer QH states
as a function of the tilt angle. The gaps at the even-integer
fillings follow qualitatively the independent-electron picture,
while the odd-integer states show rapid rise towards the coincidence
angles. For all the samples we studied, the $\nu$ = 3 valley
splitting on both sides of the coincidence shows linear density
dependence with significantly different slopes. The difference of
the $\Delta_3$(N=0,$\downarrow$) and $\Delta_3$(N=1,$\uparrow$)
gaps, as well as the observed spin-dependent resistivity, can be
qualitatively explained by screening of the Coulomb interaction from
the low-lying filled levels.

This work is supported by the NSF, the DOE and the AFOSR and the AFOSR 
contract number is FA9550-04-1-0370. We thank Dr. Qi Xiang of AMD for 
supplying us with the high quality relaxed SiGe substrates. Sandia National 
Labs is operated by Sandia Corporation, a Lockheed Martin Company, for
the DOE. The experiment performed in NHMFL is under the project
number 3007-081. We thank E. Palm, T. Murphy, G. Jones, S. Hannahs
and B. Brandt for their assistances and Y. Chen and D. Novikov for
illuminating discussions.

\end{document}